\documentstyle[twoside,fleqn,espcrc2]{article}

\newcommand{\AmS}{{\protect\the\textfont2
  A\kern-.1667em\lower.5ex\hbox{M}\kern-.125emS}}

\title{QCD correlation functions and instantons}

\author{E.Shuryak
\address{Physics Department, State University of New York,
Stony Brook NY 11794, USA}%
\thanks{Supported by the DOE
         research grants    DE-FG02-88ER40388  and
                DE-FG02-93ER40768}
}
\begin{document}

\begin{abstract}
 QCD point-to-point correlation
functions at distances .2-1 fm are very different for different channels, and
they tell us a lot about inter-quark interactions. Recent studies based on
experimental data, 'instanton liquid' approach and lattice measurements
are reviewed. Agreement between all of them show that
instanton-induced forces dominate the light quark physics, and new findings by
Negele et al, that {\it hadrons survive 'cooling'}, make this statement
obvious. We also argue that {\it chiral symmetery restoration}
is due to breaking of the 'instanton liquid'
into {\it polarized} 'instanton molecules'.
\end{abstract}

\maketitle

\section{INTRODUCTION}

  Although space is limited, let me start
with a very general comments.
 There is no need to repeat at this conference how
complicated is the QCD vacuum, and how basic it is for the whole domain of
 particle and nuclear physics. However, it seems
 worth repeating that lattice simulations should not
 only try to reproduce the
  Particle Data
Table, with the best accuracy possible, they should
lead to  {\it understanding}
of its structure.

  How one can do it? Of course, there are multiple examples
in other areas.
Just the preceeding talk gives an example of similar attemts:
cosmologists try hard to find 'large scale structures' in positions of
 galaxies and inhomogeneous background radiation. Much more familiar
 example is meteorology:  very complicated charts of
wind velocity, pressure and temperature measured
in many points can be translated to
easily undestandable forecasts, just by
pointing out
  few main cyclons, their
 magnitude and direction of motion.

  Similarly, one  may look for 'structures' among variuos fluctuations
of the gauge and quark fields, trying to single out
the most important ones.
I  am going to argue below, that it is   the 4-dimensional relatives
of cyclons, the instantons, which one should look for first.

  My second remark deals with the tools used,
the {\it point-to-point} correlation functions.
We need them because for light quark channels
    (unlike the heavy quarkonia)  the
knowledge of hadronic masses does not provide simple and unique picture of
interquark interactions. (Note
here analogy with nuclear
physics: compare 30's, when only binding of nuclei were known, to the days
after precise NN scattering experiments were done.)

 Lattice simulations  have an important role here:
  new mass measurements should be
supplimented by new wide scale studies of details unavailable
in 'real' experiments, such as correlators, wave functions and
 structure functions
 for  channels with various quantum numbers.

  Of particular interest is a
 transition region, $x\sim 0.2-0.5 fm$
 where  free  propagation of quarks and gluons
(at small distances) turns into  complicated
non-perturbative behaviour.
Discussion of relevent physical questions and available
phenomenological information on this region can be found
in my recent review \cite{Shuryak_cor}.

    Tunneling phenomena in gauge theories, discovered
in  \cite{Belavin_etal_1975}, fascinating
semiclassical theory, explanation of chiral anomalies
 \cite{tHooft},
first applications to QCD problems \cite{CALLAN_DASHEN_GROSS_1978}
etc. have attracted a lot of attention in late 70's.
 However, as no explanation
for 'diluteness'  and  validity
of semiclassical approximation
were suggested from the first principles, optimism has soon died out and
 most people  left the field.

  Next difficult period has mainly focused on
 phenomenological manifestations of instanton-induced effects.
 The so called  'instanton   liquid' model
 \cite{Shuryak_1982} has emerged as a qualitative picture. It
suggested relative diluteness and large action per instanton due to
their relatively small size, but also emphasized significant interaction
in the ensemble as the origin of density stabilization.
It was
shown,
 that this picture
 can explain several puzzles of the hadronic world
(see e.g. \cite{physrep2,Shuryak_cor} for details).

   Attempts to describe {\it interacting} instantons
were initiated by the {\it variational approach}
 \cite{DIAKONOV_PETROV_1984},   which
has qualitatively
  reproduced the 'instanton liquid' picture.
Further numerical studies of this problem \cite{Shuryak_1988}
have allowed to get rid of many approximations and
eventually included fermionic effects
to all orders in {\it 't Hooft effective Lagrangian} \cite{tHooft}.
We return to  discussion of this approach below, and now let me jump directly
to several important steps made during the last
year.

  First of all,
  more than 40 correlation
functions were calculated  in the framework of the simplest
ensemble of the kind, the Random Instanton Liquid Model (RILM)
 \cite{Shuryak_Verbaarschot_cor}. Agreement with data is
generally good, and
in some cases (including $\pi$,N etc) it is really surprising.

  Next,  some of those functions
were calculated on the lattice \cite{Negele_etal}, also with good
agreement with the RILM results.
  That the agreement is not ocasional but based on the same dynamics
 was clearly shown by recent findings, a kind of decisive experiment,
    reported at this conference \cite{Negele_DALLAS}.
'Cooled' lattice configurations, containing {\it only} instantons, not only
 has  reproduced  parameters of the
'instanton liquid', but they also lead to about the same correlation functions!

  I think these studies, taken together, has essentially  answered many
old questions like: why a nucleon is bound?
Instanton-induced attraction rather than confinement
or perturbative effects  clearly play the major role here.

\section{WHY INSTANTONS?}

  The main formal reason why instantons are so important for
physics of {\it light fermions} is related to the famous 't Hooft
'zero modes', the localized solutions of the Dirac equation
$$  D_\mu\gamma_\mu \phi_0(x)=0 $$
in the instanton field. Evaluating
the (Euclidean)  quark propagator
$ S= - 1 /[ iD_\mu\gamma_\mu +im] $
for $m \rightarrow 0$ one has to deal mainly with  small eigenvalues.

  It is convenient to look at the instanton as a 'trap' for quarks, something
 like a 'receptor' atom in
a semiconductor,  creating a new state with the energy
value being forbidden otherwize.
 At finite density of such atoms,
an electron
can propagate far, just by hopping from one atom to another. The same is
true for finite instanton density, leading to a 'zero mode zone'
 of collectivized quark states. That is the
 mechanism leading to the non-zero quark condensate.
   If more than one quark is travelling
in the
 QCD vacuum ($\bar q q$ for mesons and qqq for baryons), they  'hop'
over the same instantons. It implies an attractive interaction, which is
 in fact the one binding quarks together.

   Why instantons
and not any other fluctuation
of the gauge field? Here one has to look more specifically into
the chiral and flavor structure of the instanton-induced interaction.
As shown by 't Hooft,
 at tunneling quarks with one chirality
 'dive into the Dirac sea' while those  with the opposite chirality
 'emerge' from it. Therefore instanton-induced forces should be
stronger in scalar and pseudoscalar channels
compared to vector or axial ones.
Looking at phenomenological correlators at small distances, one finds that it
is exactly right.

   Consider corrections to correlation functions are provided by
instantons to the free propagation at small distances. (If those corrections
are relatively small, and this is the main point, one may hope to use the
't Hooft interaction in the lowest order, which makes consideration simple.)
 There are 4 such channels for 2 flavors
and it is a simple matter to see that correction is positive (or {\it
attractive}) for $\pi,\sigma$ channels and negative (or {\it repulsive}
for $\delta,\eta'$ ones. Thus {\it the same} mechanism leads to
both light pion and heavy $\eta'$! Both splitting from 'typical
mesons' are  large, which is
a very strong hint.

\section{CORRELATORS IN THE INSTANTON VACUUM}

  Now we proceed from  qualitative hints to quantitative calculations.
We have to
 evaluate
 a quark propagator in the multi-instanton
field configuration, which can be done
 as follows:
$$ S(x,y)=\Sigma_{ZMZ}
 {\phi_\lambda(x) \phi^+_\lambda(y) \over \lambda -im }+ iS_{NZM}(x,y))
$$
where the first term is the sum over states belonging to the 'zero mode zone'.
The non-zero modes
(analogs of 'scattering states') are taken into account by the last term, see
details in
\cite{Shuryak_Verbaarschot_cor}.

  We have first calculated correlators for the
  simplest ensemble possible, the {\it 'random instanton liquid model'}
 (RILM), in which: (i) all instantons
have the same size $\rho_0=.35 fm$; (ii) they have {\it random}
 positions and orientations;
(iii) instanton and anti-instanton densities are equal, and in sum it is
$n_0=1 fm^{-4}$.
  These are the parameters suggested a decade ago in \cite{Shuryak_1982}, the
density comes from
the gluon condensate and size from various other things, say from
the quark condensate value. 'Cooling' of lattice configurations provide
a way to see instantons and check them, see \cite{Negele_DALLAS} for the
latest results and earlier references. Let me only state that these numbers
are essentially confirmed.

The main step in the calculation is inversion of the Dirac
operator, written in the zero-mode subspace. (We typically
use in sum 256 instantons and
 anti-instantons, which tells the dimension of this matrix and the volume of
the
box.)

   Although the quark propagators are gauge dependent, we have looked
at them first in order to see whether
they can be reproduced by any simple model, say by 'constituent quarks' with
a constant mass. We have found that chirality-non-flipping part of the
propagator
indeed looks as if quark get a mass about 300-400 MeV, but the
 chirality-flipping part  does not look like that
at all. None of many correlators calculated behave like 'constituent quark
model'.

\begin{figure}[htb]
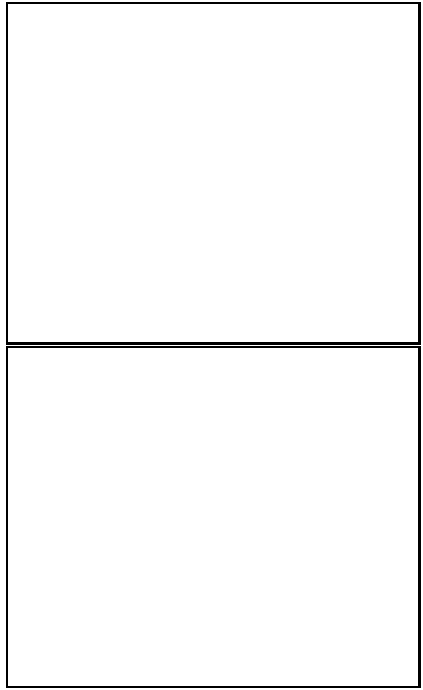

\vspace{9pt}
\framebox[55mm]{\rule[-21mm]{0mm}{43mm}}
\framebox[55mm]{\rule[-21mm]{0mm}{43mm}}
\caption{ Correlators for $\pi,\rho$ channels according to
RILM results (the closed triangles) and lattice calculation
(open squares) versus distance x in fm.
}
\label{fig:RILM}
\end{figure}
\begin{figure}[htb]
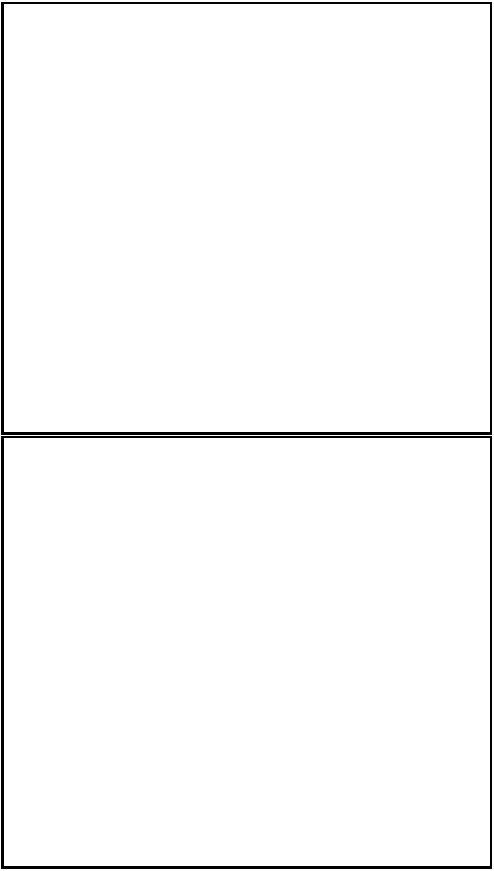

\vspace{9pt}
\framebox[65mm]{\rule[-21mm]{0mm}{55mm}}
\framebox[65mm]{\rule[-21mm]{0mm}{55mm}}
\caption{ Same as in Fig.1 but for N,$\Delta$ channels}
\label{fig:RILM2}
\end{figure}

   Our results for $\pi,\rho,N,\Delta$ channels are shown in
Fig.\ref{fig:RILM} and Fig.\ref{fig:RILM2}
 All correlators are plotted in a normalized way, divided by
 those correcponding to {\it free} quark propagation:
that is why all of them converge to 1 at small distances.
Solid lines correspond to experiment \cite{Shuryak_cor},
while the  long-dashed and short-dashed curves correspond to
QCD sum rule predictions
\cite{Belyaev_Ioffe} and \cite{FZOZ}, respectively.

The agreement for the pion curve is as  perfect as it can be:
 both the mass (142$\pm$12 MeV) and the (pseudoscalar)
  coupling are reproduced correctly, inside
the error bars! Note large deviations from perturbative behavior
happens at very small distances for the $\pi$ channel, while exactly the
opposite is observed in the $\rho$ case,
the plotted ratio
remains close to 1 up to a very large x.
Both RILM and lattice has reproduced that non-trivial observation.

  Proceeding to baryonic channels, let me mention that
 we have actually
measured all 6 nucleon correlators and 4 delta ones, and have fitted them all.
  Again, agreement between RILM and lattice results is surprisingly good,
literally inside the error bars.
Both display a qualitative difference
between the nucleon and the delta correlators: this can be
traced to {\it attractive}
 instanton-induces forces for the spin-isospin-zero diquarks
\cite{Betman_Laperashvili}.
Without any one-gluon exchange the RILM
predicts the $N-\Delta$ splitting (actually,
we have found that in RILM $m_N=960\pm 30 MeV$ and  $m_\Delta= 1440\pm70 MeV$,
so the splitting is
in fact somewhat too large).

  There is  no place here to discuss many other channels.
The most difficult case proved to be
 the isosinglet scalar $\sigma$, for which
 one should not only evaluate
the 'double quark loop' term, but also subtract the disconnected
$|<\bar q q>|^2$ part. Curiously enough, we have found dominance of
a light state, with
m$\sim$ 500 MeV, reminiscent of the 'sigma meson' of 60's.
For several  reasons such measurements are now beyond the reach
of lattice calculations, although existence of attractive interaction
at $x\sim 1/2 fm$ can probably be seen.

  Not included in the preprints mentiond are studies of the
 the so called 'wave functions' (known also as {\it Bethe-Salpeter amlitudes}).
They have shown with greater clarity that RILM
does lead to  quark binding, without
perturbative (Coulomb-like) and confining forces.
Although the shape of the wave function is not the same as was
found on the lattice, its width is only slightly larger.
The main qualitative features (e.g.: $\pi,N$ are {\it more compact} than
$\rho,\Delta$) are also reproduced.

\section{INTERACTING INSTANTONS}

  Clearly, RILM discussed above cannot be but a crude
approximation:
 at least, the very phenomenon studied
above,  'hopping' of quarks from one instanton to another, should
 produce strong
correlation between them. Another obvious source of interaction is the
non-linear gluonic Lagrangian: a superposition of instantons and
anti-instantons
 have the
action different from the sum of the actions.
Effective
  statistical system should be described by a partition function
$$ Z= \int  d\Omega  exp(-S_{glue}) [det(i\hat D
+im)]^{N_f}
$$
(where
$ d\Omega$
 is  the measure
in space of {\it collective coordinates}, 12 per instanton).
It is
a problem similar to those traditionally studied in statistical mechanics,
the compication is the fermionic
determinant which is non-local. However,  it is
still orders and orders of magnitude simpler than the lattice gauge theory.
  As
shown in
\cite{DIAKONOV_PETROV_1984,Shuryak_1988}
this statistical sum describes a liquid, in which
chiral symmetry {\it is broken}.
   Recent studies of correlation functions with {\it interacting ensemble}
have shown other
significant improvement over RILM. In particular, the global
fluctuations of the topological charge are screened, and
for $m\rightarrow 0$ the topological
susceptibility vanishes, as it should. The distribution of Dirac eigenvalues
$\lambda$ has a different shape, with a dip at $\lambda=0$ instaed of a peak,
 and  satisfy several non-trivial general theorems, see
\cite{Verbaarschot_Zakopane}.

   Unfortunately, it is still not quite quantitative theory. The reason
is we do not know how to separate 'semiclassical' fluctuations with
large action, from very close instanton-anti-instanton pair
on the bottom of the 'valley'
(\cite{Yung}).
Those configurations has small field, therefore
they should be included in the
perturbation theory. If we suppress them by some {\it repulsive core}, we get
reasonable results, but a satisfactory
 solution of this problem is still missing.

\section{CHIRAL SYMMETRY RESTORATION}
  Instantons at finite temperatures
is a subject for separate talk, and here let me only discuss recent
development related with critical phenomena near the
 chiral symmetry restoration
point, $T\approx T_c$.

 The main phenomenon in this region is
 a strong 'pairing' of instantons, leading to
splitting of the 'instanton liquid' into the $\bar I I$ molecules.
The first (strongly simplified) discussion of chiral restoration
transition
at this angle was made in \cite{IS}.

\begin{figure}[htb]
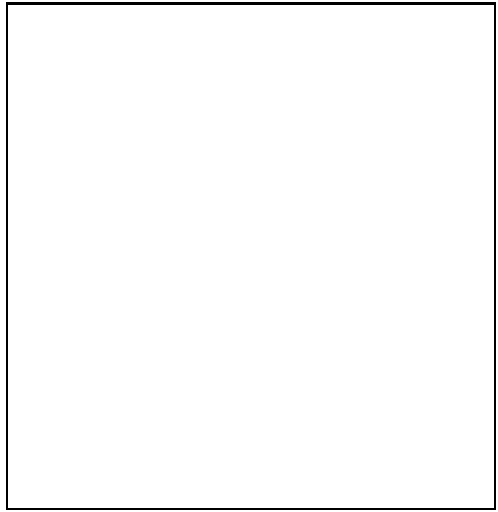

\framebox[65mm]{\rule[-21mm]{0mm}{65mm}}
\caption{Fraction of the $\bar I I$ molecules for which $\tau > r$
 for QCD (squares) and
fir 4 massless quark flavors (stars).
}
\label{fig:polarization}
\end{figure}

   One new finding is strong and rapid 'polarization' of these molecules
in the critical region. $\bar I I$ interaction at
finite T were studied in details in \cite{SV_T}, and the main anisotropy
comes from the quark-induced interaction
  $\sim |sin(\pi T \tau)/cosh(\pi T r)|^{2N_f}$,
(where $\tau,r$ are distance between the two centers in time and space
directions). Around $T-c$ the point $r=0,\tau=1/(2T)$ is strongly prefered,
and, as shown in Fig.\ref{fig:polarization}, the
 degree of polarization rapidly frows.

  This polarization leads to rapid growth of the
gluonic energy density
in the transition region, which was
in fact observed in lattice simulations with
dynamical fermions.
 Recall that, in terms of Minkowski field strengths, it is
$$ \epsilon={1\over 2}(E^2+B^2)+ g^2{(11/3)N_c-(2/3)N_f \over 128\pi^2}
(E^2-B^2)
$$
For 'unpolarized' objects  the first term is zero,
and only the second ('anomalous') one contributes. However, for the
 'polarized' molecules one finds $E^2=-.8 B^2$, and the first term works. It
produces a
jump in $\epsilon(T)$ of the order of 1 $ GeV/fm^3$.

\begin{figure}[htb]
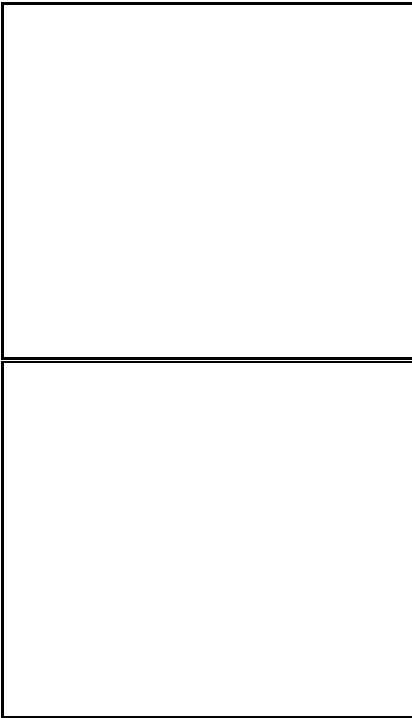

\framebox[55mm]{\rule[-21mm]{0mm}{45mm}}
\framebox[55mm]{\rule[-21mm]{0mm}{45mm}}
\caption{$\pi,\rho$ correlation functions for the 'cocktail' model.
Open points correspond to pure 'random' instanton ensemble, while
the closed ones correspond to $f_m=1$. The lines show intermediate
cases $f_m= .25,.5,.75$.
}
\label{fig:cocktail}
\end{figure}

   The second interesting findings are
 specific changes in the correlation
functions below $T_c$. A simple
'cocktail model'
  \cite{SSV_mix}  was used,
containing both 'random' component and some
'molecule fraction' with weight $f_m=2 N_{molecules}/ N_{all}$.
We have first found that $<\bar q q>$,
depends  on $f_m$ in a way similar to
its T-dependence, measured on the lattice: it changes little first, and then
rapidly vanishes at $f_m\rightarrow 1$.
Different correlation functions depend
on $f_m$ quite differently. Two examples are shown in Fig.\ref{fig:cocktail}
  for $\pi,\rho$
channels.
One can see, that
while the vector correlator is not changed much, the pion on display
remarkable stability for $f_m=0-0.8$, with subsequent strong drop
toward $f_m=1$. At the last point {\it complete}
 chiral symmetry get restored, so the pion correlator
 coinsides with
its scalar partners $\sigma,\delta$.
More studies along these lines and their comparison with recent
lattice works are in progress.

\end{document}